\documentclass[letter]{aa} % for the letters 
\usepackage{graphicx}
\usepackage{txfonts}
\newcommand{\okina}{`}
\usepackage[normalem]{ulem}

\begin{document}

	\title{2I/Borisov: A C$_2$depleted interstellar comet}
	
	%   \subtitle{I. Overviewing the $\kappa$-mechanism}

	\author{Cyrielle Opitom\inst{1,2}
		\and
		Alan Fitzsimmons\inst{3}
		\and
		Emmanuel Jehin\inst{4}
		\and
		Youssef Moulane\inst{1,4,5}
		\and
		Olivier Hainaut\inst{6}
		\and
		Karen J. Meech\inst{7}
		\and
		Bin Yang\inst{1}
		\and
		Colin Snodgrass\inst{2}
		\and
		Marco Micheli\inst{8,9}
		\and
		Jacqueline V. Keane\inst{7}
		\and
		Zouhair Benkhaldoun\inst{5}
		\and
		Jan T. Kleyna\inst{7}
	}
	
	\institute{European Southern Observatory, Alonso de Cordova 3107, Vitacura, Santiago, Chile 
		\email{copitom@eso.org}
		\and
		Institute for Astronomy, University of Edinburgh, Royal Observatory, Edinburgh EH9 3HJ, UK
		\and 
		Astrophysics Research Centre, Queen's University Belfast, Belfast BT7 1NN, UK 
		\and
		STAR Institute, Universit\'e de Li\`ege, All\'ee du 6 aout, 19C, 4000 Li\`ege, Belgium
		\and 
		Oukaimeden Observatory, Cadi Ayyad University, Morocco
		\and
		European Southern Observatory, Karl-Schwarzschild-Strasse 2, D-85748 Garching bei M\"unchen, Germany
		\and 
		Institute for Astronomy, 2680 Woodlawn Drive, Honolulu, HI 96822 USA
		\and
		ESA NEO Coordination Centre, Largo Galileo Galilei, 1, 00044 Frascati (RM), Italy
		\and
		INAF - Osservatorio Astronomico di Roma, Via Frascati, 33, 00040 Monte Porzio Catone (RM), Italy 
	}

	\date{Received October xx, 2019; accepted October xx, 2019}
	
	% \abstract{}{}{}{}{} 
	% 5 {} token are mandatory
	
	\abstract
	{}
	{The discovery of the first active interstellar object 2I/Borisov provides an unprecedented opportunity to study planetary formation processes in another planetary system. In particular, spectroscopic observations of 2I allow us to constrain the composition of its nuclear ices. }
	{We obtained optical spectra of 2I with the 4.2~m William Herschel and 2.5~m Isaac Newton telescopes between 2019 September 30 and October 13, when the comet was between 2.5~au and 2.4~au from the Sun. We also imaged the comet with broadband filters on 15 nights from September 11 to October 17, as well as with a CN narrow-band filter on October 18 and 20, with the TRAPPIST-North telescope.}
	{Broadband imaging confirms that the dust coma colours (B-V=0.82$\pm$0.02, V-R=0.46$\pm$0.03, R-I=0.44$\pm$0.03, \mbox{B-R=1.28$\pm$0.03}) are the same as for Solar System comets. We detect CN emission in all spectra and in the TRAPPIST narrow-band images with production rates between $1.6\times10^{24}$ and $2.1\times10^{24}$ molec/s. No other species are detected. We determine three-sigma upper limits for C$_2$, C$_3$, and OH production rates of $6\times10^{23}$ molec/s, $2\times10^{23}$ molec/s and $2\times10^{27}$ molec/s, respectively, on October 02. There is no significant increase of the CN production rate or A(0)f$\rho$ during our observing period. Finally, we place a three-sigma upper limit on the Q(C$_2$)/Q(CN) ratio of 0.3 (on October 13). From this, we conclude that 2I is highly depleted in C$_2$, and may have a composition similar to Solar System carbon-chain depleted comets. }
	{}
	
	\keywords{comets: individual (2I/Borisov) ---  comets: general}
	
	\maketitle
	%
	%-------------------------------------------------------------------
	
	\section{Introduction}
	
	Ejected planetesimals are expected as a result of giant planet migration in planetary systems \citep{Levison2010,Dones2004}. Consequently, it has been foreseen for decades that interstellar objects (ISOs) should cross the Solar System or even be captured, providing opportunities to study planetary formation processes in other systems \citep{engelhardt2017}. The first ISO, 1I/2017 U1 (\okina Oumuamua), was discovered on 2017 October 17 at 1.12~au from the Sun, post-perihelion \citep{meech2017}. Even though non-gravitational acceleration of its motion suggested 1I was active \citep{micheli2018}, no gas or dust coma was detected \citep{meech2017,fitzsimmons2018,ye2017,trilling2018}. The lack of coma detection around 1I unfortunately prevented any detailed study of the composition of the first interstellar object.
	
	On 2019 August 30, comet C/2019 Q4 (Borisov) was discovered by amateur astronomer Gennady Borisov. The comet was quickly found to have a hyperbolic orbit, with an eccentricity $>3$, and was officially renamed 2I/Borisov (hereafter 2I) on September 24. Contrary to 1I, 2I was observed to be active  at the time of discovery at 2.985~au from the Sun. Initial spectroscopic observations by \cite{de_Leon_2019} revealed a featureless spectrum. Optical dust colours were found to be consistent with those of active Solar System comets \citep{fitzsimmons2019,guzik2019,jewitt2019}. The  radius of the nucleus was estimated to be between 0.7~km and 3.3~km by \cite{fitzsimmons2019}, consistent with the upper limit of 3.8~km set by \cite{jewitt2019}. Observations of 2I by \cite{fitzsimmons2019} allowed the first detection of gas around an interstellar object. From the detection of the CN (0-0) band at 388nm, the CN production rate of 2I was estimated to be of $3.7\times10^{24}$ molec/s on September 20 at 2.66 au. Recently, \cite{kareta2019} claimed a detection of C$_2$ with a production rate of $2.5\times10^{24}$ molec/s. %, indicating that the first interstellar comet could be slightly depleted in C$_2$.
	
	In this letter, we present spectroscopic and photometric observations of 2I aimed at constraining its composition, complemented with broadband imaging to monitor the comet activity and dust colours. 
	
	\section{Observations and data reduction} 
	
	\label{sec:observations}
	\subsection{William Herschel Telescope}
	2I/Borisov was observed with the 4.2~m William Herschel Telescope (WHT) and the ISIS spectrograph on La Palma on 2019 October 02.2 UT and 13.2 UT. The observational circumstances are given in Table~\ref{tab-obslog}. The spectrograph and detector configuration was as described in \cite{fitzsimmons2019}, except that the centre of the observed spectral range was moved to 406.6~nm. This allowed observation of wavelengths from the atmospheric cut-off at 300~nm up to 580~nm \ within the unvignetted field of the spectrograph camera, covering many prominent molecular emission bands observed in normal comets. On both nights we obtained three consecutive 1200 second exposures, but on October 02.2 UT the last one was unusable due to bright twilight sky. On both dates the comet exposures were immediately followed by exposures of the spectrophotometric standard G191-B2B \citep{bohlin1995} using a 10\arcsec-wide slit to enable flux calibration, and the solar analogue HD~28099 \citep{hardorp1980} using the same 2\arcsec-wide  slit as used for the comet to enable removal of the contribution of the dust-reflected continum.  The comet spectra were extracted over an aperture of 8\arcsec\ centred on the comet.
	
	\subsection{Isaac Newton Telescope}
	Observations of 2I were also performed with the 2.54~m Isaac Newton Telescope combined with the IDS spectrograph on La Palma on 2019 September 30.2 UT and October 01.2 UT.  The IDS is a low-resolution spectrograph with an unvignetted slit length of 3.3\arcmin. We used the blue-sensitive EEV10 detector and the IDS R400B grating providing a dispersion of 0.141~nm/pixel at 390~nm, combined with a 2\arcsec-wide slit.  We obtained three 1200~s exposures on the first night and four 1200~s exposures on the second night. However, the last exposure of each night was affected by strong background of the morning twilight. As such, we decided to discard those exposures.  The spectra were bias subtracted, flat fielded, and wavelength calibrated using a CuAr+NeAr lamp. The spectrophotometric standard star HILT600 was observed immediately before 2I on each night and was used to flux calibrate the spectra. The spectra were extracted over the same 8\arcsec\ aperture as for the WHT, and all five spectra were then co-added to increase the signal-to-noise ratio. Finally, the contribution of the dust-reflected continuum was removed using the scaled and reddened spectrum of a solar analogue. 
	
	\subsection{TRAPPIST}
	The interstellar comet 2I was also observed with the 60~cm TRAPPIST-North telescope (TN) located at Oukaimeden observatory, Morocco \citep{Jehin2011}. TN is equipped with a 2k$\times$2k CCD camera with a field of view of 22$\arcmin \times$22$\arcmin$. We binned the pixels 2 by 2 and obtained a resulting plate scale of 1.2$\arcsec$/pixel. We observed 2I with broad-band BVRI filters \citep{Bessell1990} on 15 nights from 2019 September 11, when the comet was at 2.80~au from the Sun, to October 17 (r$_h$=2.31~au).  On each night, three images were obtained  in the B,V, and I filters and about six images were taken in the R filter, all with an exposure time of 180~s.
	%We used the to measure the magnitudes, colours, and the Af$\rho$ parameter \citep{AHearn1984afp} of 2I.  
	On October 18 and 20, we also obtained an exposure with the CN narrow-band filter \citep{Farnham2000}, with an exposure time of 1500~s. Data reduction followed standard procedures using frequently updated master bias, flat, and dark frames. The sky contamination was subtracted and the flux calibration was performed using regularly updated zero points based on observations of photometric standard stars,  as described in previous papers \citep{opitom2016,Moulane2018}.
	
	\section{Analysis}
	\label{sec:analysis}
	\subsection{The gas coma}
	
	\begin{figure*}[h!]
		\begin{center}
			\includegraphics[scale=0.60]{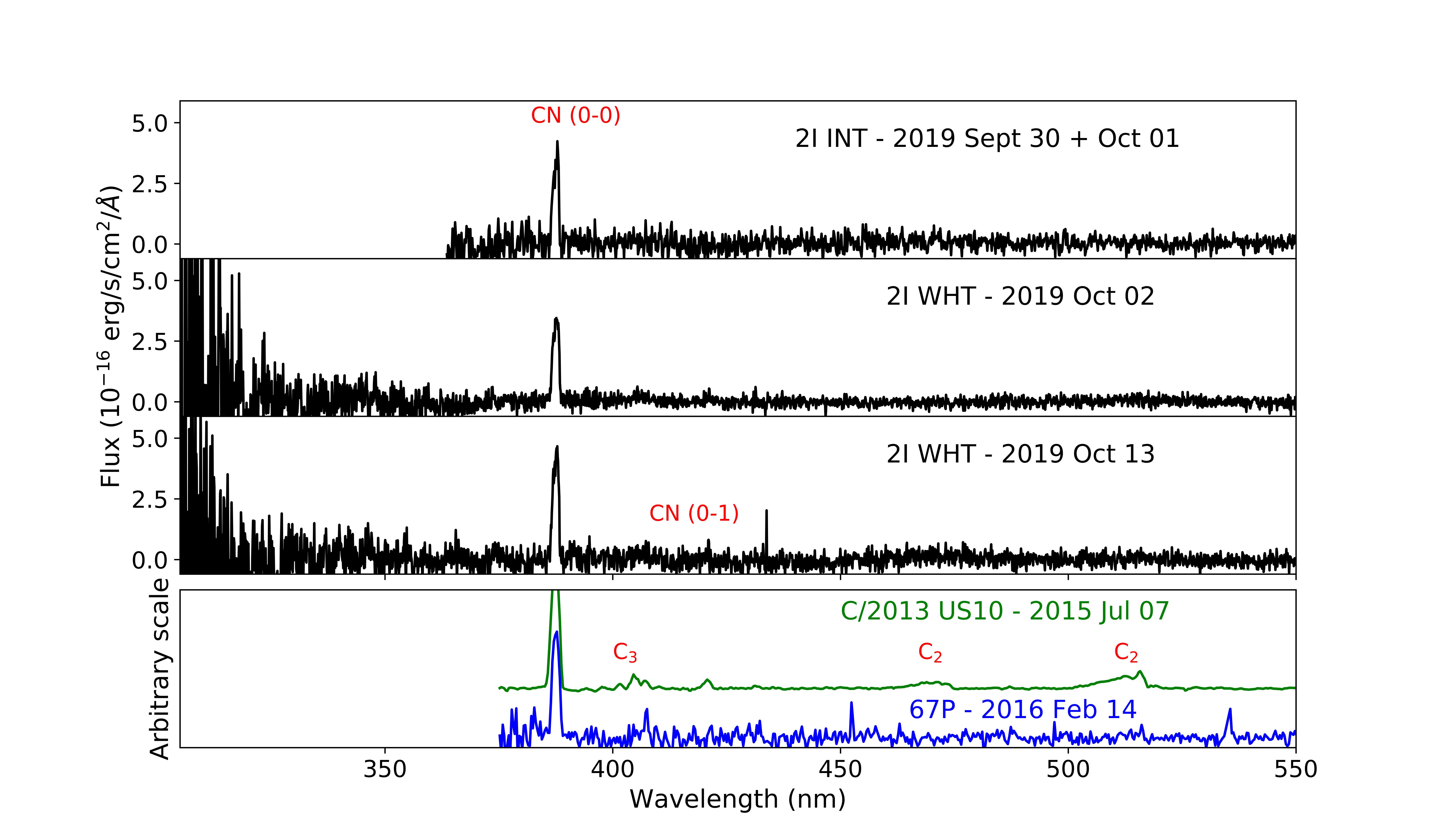}
			\caption{Flux calibrated and continuum subtracted spectra of 2I/Borisov obtained with the INT and WHT through a 2\arcsec$\times$8\arcsec\ aperture centred on the comet. The spike around 435~nm in the October 13 spectrum is a noise feature. The bottom plot shows spectra of two comets observed at similar heliocentric distances as 2I: 67P/Churyumov-Gerasimenko (from \cite{opitom2017}, r$_h=2.36$~au), a Jupiter Family comet with a CN production rate similar as 2I and for which C$_2$ was not detected at that distance, and C/2013 US10 (Catalina) (from \cite{opitomthesis}, r$_h=2.30$~au), a typical very active long-period comet, for which both C$_2$ and C$_3$ were detected. Those spectra have both been arbitrarily scaled on the y-axis.}
			\label{fig:Spectra}
		\end{center}
	\end{figure*}
	
	\begin{figure}[h!]
		\begin{center}
			\includegraphics[scale=0.45]{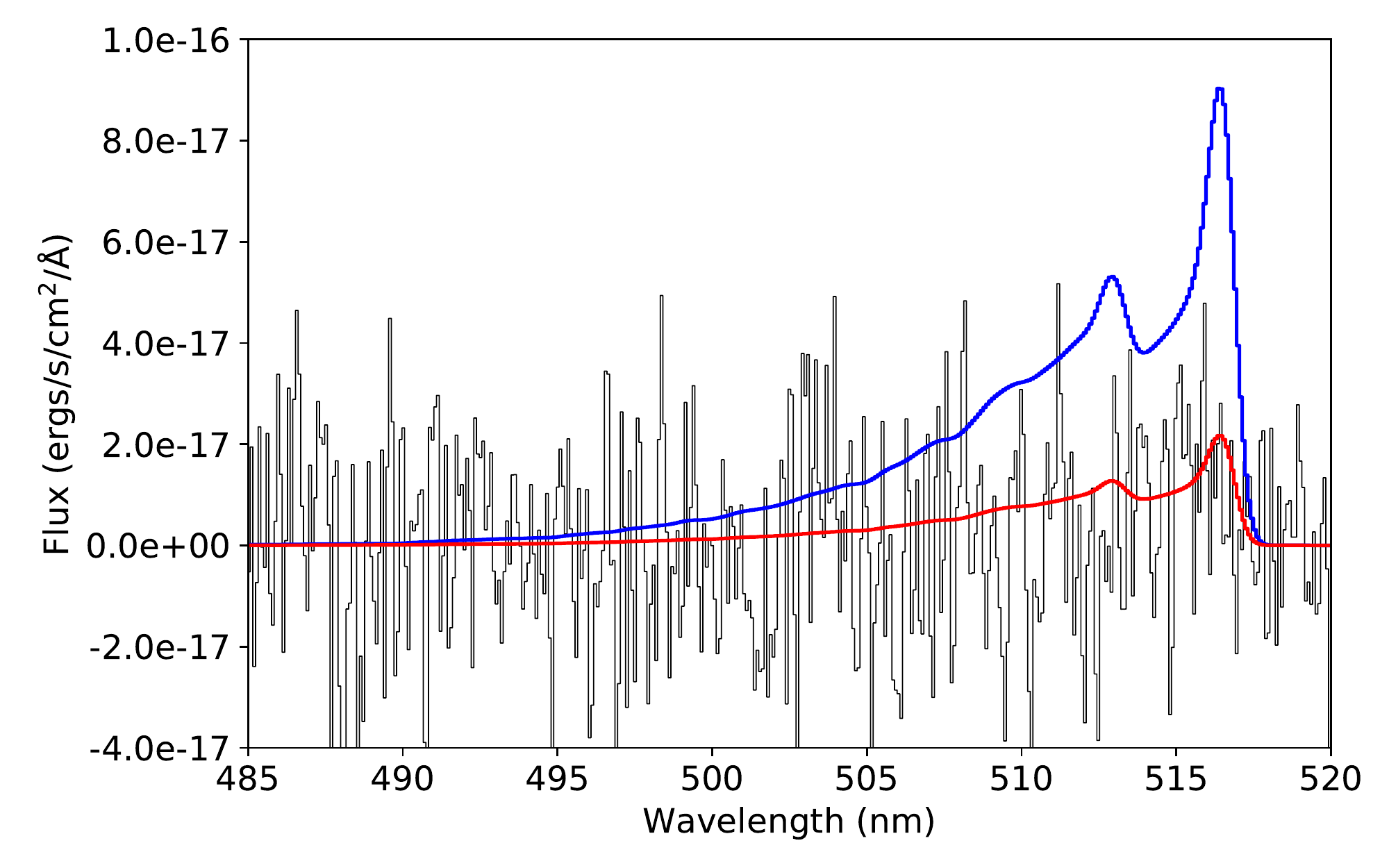}
			\caption{Black data: Continuum-subtracted spectrum of 2I/Borisov on 2019 October 13.2 UT  in the region of the C$_2$(0-0) emission band. Blue theoretical spectrum: C$_2$ emission flux through our aperture for a production rate of Q(C$_2$)$=2.5\times10^{24}$ molec/s at a spectral resolution of $\Delta \lambda=1$nm . Red theoretical spectrum: as above for Q(C$_2$)$=6\times10^{23}$ molec/s.}
			\label{fig:C2spectrum}
		\end{center}
	\end{figure}
	
	\begin{table*}[h!]
		\begin{center}
			\caption{Log of observations and production rates of 2I/Borisov with the WHT, INT,  and TRAPPIST-North. }
			\label{tab-obslog}
			\begin{tabular}{lccccccccc}
				\hline
				Date &  Telescope & r$_h$ & Delta & Exp. & Exp Time & Q(CN) & Q(C$_2$) & Q(C$_3$)  & Q(OH) \\
				UT &  & au & au &  & s & molec/s & molec/s & molec/s & molec/s  \\
				\hline
				\hline 
				Sep 30-Oct 01  & INT & 2.51 & 3.01 & 5 & 1200 & $(1.8\pm0.1)\times10^{24}$ & $<9\times10^{23}$ & $<3\times10^{23}$ & --\\
				Oct 02  & WHT & 2.50 & 3.00 & 2 & 1200 & $(1.9\pm0.1)\times10^{24}$ & $<6\times10^{23}$ & $<2\times10^{23}$ & $<2\times10^{27}$\\
				Oct 13  & WHT & 2.36 & 2.77 & 3 & 1200  & $(2.1\pm0.1)\times10^{24}$ & $<6\times10^{23}$ & $<3\times10^{23}$ & $<2\times10^{27}$\\
				Oct 18  & TN & 2.31 & 2.67 & 1 & 1500  & $(1.9\pm0.6)\times10^{24}$ & - & - & -\\
				Oct 20  & TN & 2.29 & 2.64 & 1 & 1500  & $(1.6\pm0.5)\times10^{24}$ & - & - & -\\
				\hline
			\end{tabular}
		\end{center}
	\end{table*}
	
	In all reduced spectra, the CN~(0-0) emission at 388~nm is clearly detected, as can be seen in Fig. \ref{fig:Spectra} where the three spectra of 2I are compared to the spectra of two Solar System comets observed at a similar heliocentric distance. The flux was measured by approximating the band shape with two free-fitted Gaussians. We used the fluorescence scattering efficiency from \cite{schleicher2010} to compute the total number of CN molecules within the extraction aperture. We then used a Haser model \citep{haser1957} to compute the CN production rates for all our observing dates, with an outflow velocity of $0.85/\sqrt{r_h} \simeq 0.5$ km s$^{-1}$ \citep{cochran1993} and effective scale lengths from \cite{ahearn1995}. In order to derive the CN production rate from TRAPPIST narrow-band images, we derived median radial brightness profiles for the CN and R images and used the continuum filter to remove the dust contamination. We then converted the flux to column density and fitted the profile with a Haser model at 10,000~km from the nucleus, using the same parameters as for the spectra. In the last spectrum, we also marginally detect the CN $(0-1)$ band around 420~nm (see Fig. \ref{fig:Spectra}). The flux ratio was measured as $(0-1)/(0-0)=0.05\pm0.01$, slightly smaller than the expected value of 0.08 from the respective g-factors.

	We carefully searched for the C$_2$ (0-0) emission band around 516.7~nm, which is usually the second species detected in the optical spectrum of comets, but without success. No emission from OH at 308~nm or C$_3$ around 405~nm could be detected either. There is some suggestion of excess emission near the wavelengths of the C$_3$ band, but it does not match the expected profiles of either C$_3$, CO$^+$, or CO$_2^+$. Measuring the rms uncertainty around the wavelengths of OH, C$_2$, and C$_3$, we computed three-sigma upper limits for the production rates of those species using a Haser model and the same parameters as mentioned above. Scattering g-factors for OH~(0-0) were taken from \cite{schleicher1988}.  To support these upper limits for the C$_2$~(0-0) band, we calculated theoretical emission spectra for a cometary atmosphere at the same observational geometry as 2I using the Planetary Spectrum Generator \citep{villanueva2018}.
	In Fig. \ref{fig:C2spectrum} we overlay our data with these theoretical spectra convolved to our spectral resolution of $\sim1$nm for production rates of Q(C$_2$)$=2.5\times10^{24}$ molec/s as reported by \cite{kareta2019}, and for our upper limit of Q(C$_2$)$=6\times10^{23}$ molec/s.

	The CN production rates and the upper limits for C$_2$, C$_3$, and OH are given in Table \ref{tab-obslog}. The CN production rates we derive from the WHT and INT observations are in good agreement with TRAPPIST data within the error bars. We do not see a significant increase of the CN production rate over the three weeks between the first and last observations. The CN production rates we measure are about a factor of two lower than the one reported by \cite{fitzsimmons2019}. However, we consider them to be  consistent with that measurement considering the large uncertainty due to the high airmass and thin clouds reported by \cite{fitzsimmons2019}. The upper limit we derive for the $\mathrm{C_2}$ production rate is significantly more constraining than the one from \cite{fitzsimmons2019} ($4\times10^{24}$ molec/s). Our CN production rates are more than a factor of two lower than the CN production rate measured by \cite{kareta2019}. \cite{kareta2019} reported the detection of C$_2$ with a production rate of $2.5\times10^{24}$ molec/s, while we do not detect any C$_2$ in our observations and place a much lower upper limit on the C$_2$ production rate of $6\times10^{23}$ molec/s. Given that our more stringent upper limit was obtained at a similar time to that of \cite{kareta2019}, and that we see no evidence for significant variation in production rates between spectra, we argue that their value should also be regarded as an upper limit; careful inspection of their spectrum (Fig. 2 of \cite{kareta2019}) does not reveal any strong evidence for a detection. Indeed, a personal communication by T. Kareta on Oct. 24, 2019 confirms that the claimed C$_2$ detection in that submitted manuscript is more likely an upper limit.

	\subsection{The dust coma}
	\begin{figure}[h!]
		\begin{center}
			\includegraphics[scale=0.5]{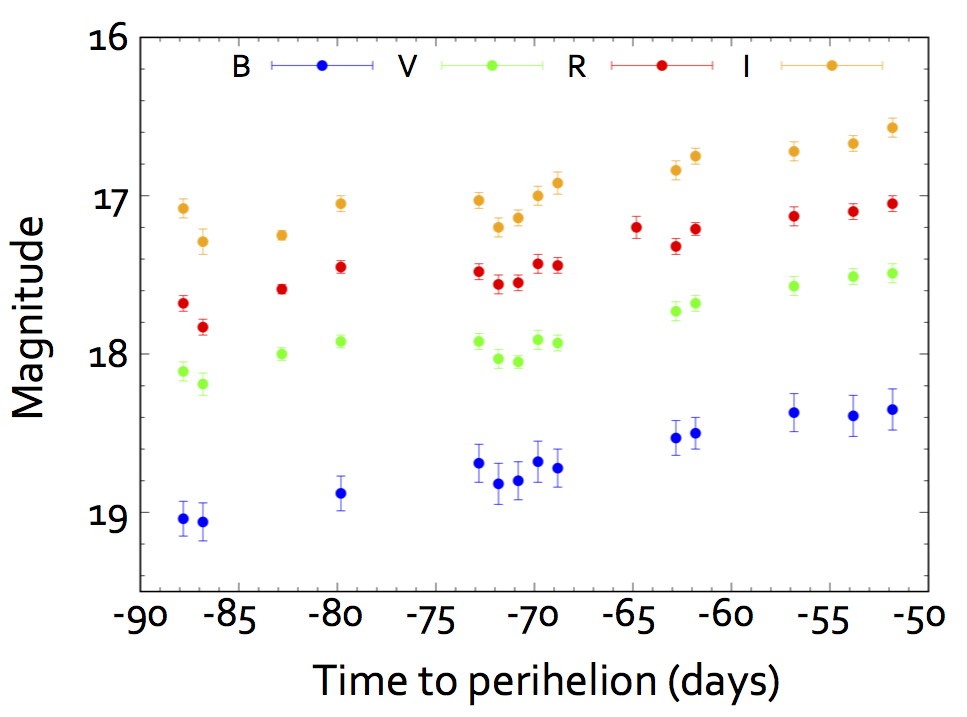}
			\caption{Evolution of the BVRI magnitude of 2I/Borisov as a function of time to perihelion.}
			\label{fig:mags}
		\end{center}
	\end{figure}
	
	We measured the magnitude of 2I in the B, V, R, and I filters within an aperture radius of 4.2$\arcsec$ corresponding to about $9\,000$~km at the distance of the comet. The brightness of 2I increased regularly but slowly in all filters, with a total increase of about 0.7 mag over one month. 
	%On October 17, we measured magnitudes in the various bands of B=18.35$\pm$0.12, V=17.49$\pm$0.06, R=17.05$\pm$0.05, and I=16.57$\pm$0.06. 
	Figure \ref{fig:mags} shows the evolution of the magnitudes as a function of time to perihelion. The average colours of the coma (dominated by the dust-reflected continuum) over the whole observing period are  given in Table \ref{tab:colours} and compared to those of 1I and to typical Jupiter Family Comets (JFCs) and Long-Period Comets (LPCs).
	%B-V=0.82$\pm$0.02, V-R=0.46$\pm$0.03, R-I=0.44$\pm$0.03, B-R=1.28$\pm$0.03. 
	The colours do not appear to vary with time. They are redder than the Sun and 1I, and similar to those measured for active JFCs and LPCs. They are in agreement with previous measurements by \cite{fitzsimmons2019,jewitt2019}.
	
	\begin{table}[ht!]
		\begin{center}
			
			\caption{Colours of 2I/Borisov compared to those of 1I/\okina Oumuamua, JFCs, and LPCs.}
			\label{tab:colours}
			\resizebox{0.48\textwidth}{!}{%
				\begin{tabular}{lccccl}
					\hline
					\hline
					Object &  \multicolumn{4}{c}{Colours}  &Ref. \\
					&     B - V   &V - R   &R - I   &B - R  & \\
					\hline
					2I/Borisov &0.82$\pm$0.02 &0.46$\pm$0.03 &0.44$\pm$0.03 &1.28$\pm$0.03 & (1)\\
					&0.80$\pm$0.05 &0.47$\pm$0.03 &0.49$\pm$0.05 &1.27$\pm$0.04 & (2) \\
					1I/\okina Oumuamua &0.70$\pm$0.06 &0.45$\pm$0.05 &-&1.15$\pm$0.05 & (3)\\
					Active LPCs &0.78$\pm$0.02 &0.47$\pm$0.02 &0.42$\pm$0.03 &1.24$\pm$0.02   & (4)\\
					Active JFCs &0.75$\pm$0.02 &0.47$\pm$0.02 &0.44$\pm$0.02 &1.22$\pm$0.02    & (5) \\
					\hline  
					Sun &0.64$\pm$0.02 &0.35$\pm$0.01 &0.33$\pm$0.01 &0.99$\pm$0.02    & (6) \\
					\hline
			\end{tabular}}
			\tablebib{ (1) This Work; (2) \cite{jewitt2019}; (3) \cite{Jewitt2017}; (4) \cite{Jewitt2015};(5) \cite{Solontoi2012}; (6) \cite{Holmberg2006}}
		\end{center}
	\end{table}
	
	We computed the Af$\rho$ parameter at $10\,000$~km from the nucleus and corrected it for the phase angle effect according to the phase function normalised at $\theta$=0$^\circ$ derived by D.~Schleicher\footnote{\url{http://asteroid.lowell.edu/comet/dustphase.html}}. We do not see a significant increase of the A(0)f$\rho$ with time or heliocentric distance and compute a mean value of A(0)f$\rho$=132.4$\pm$4.7 cm in the R filter for our data set. Using an average CN production rate of $1.9\times10^{24}$ molec/s, we derive a dust-gas ratio of log[A(0)f$\rho$/Q(CN)]$=-22.16$.
	
	\section{Discussion}
	\label{sec:discuss}
	%\subsection{Comparison with Solar System comets}
	The sensitive upper limits on the $\mathrm{C_2}$ production rate of 2I that we derive allowed us to compute an upper limit of the Q($\mathrm{C_2}$)/Q(CN) ratio in the coma of 2I of 0.3 (for October 13). With such a low Q($\mathrm{C_2}$)/Q(CN) ratio, 2I would be classified as a carbon-chain depleted comet in our Solar System (defined by log[Q($\mathrm{C_2}$)/Q(CN)]$<-0.18$ according to \cite{ahearn1995}).
	
	Numerous studies of Solar System comets have focused on the composition of the coma, including among others the $\mathrm{C_2}$, $\mathrm{C_3}$, and CN radicals. \cite{ahearn1995} showed that two main classes of comets can be distinguished based on the relative abundance of radicals observed at visible wavelengths: typical comets, and comet depleted in carbon-chain species, like $\mathrm{C_2}$ and $\mathrm{C_3}$. Later studies by \cite{fink2009,langland2001,cochran2012,schleicher2008} confirmed the existence of a class of comets depleted in carbon-chain species. \cite{ahearn1995} also showed that the fraction of depleted comets is higher among JFCs (with a source in the Kuiper Belt region) than among LPCs from the Oort cloud. This conclusion was supported by \cite{cochran2012}, who observed that 18.5\% of the LPCs in their sample were depleted, while the proportion of carbon-chain depleted JFCs was 37\%.
	
	\begin{figure*}[ht!]
		\begin{center}
			\includegraphics[scale=0.38]{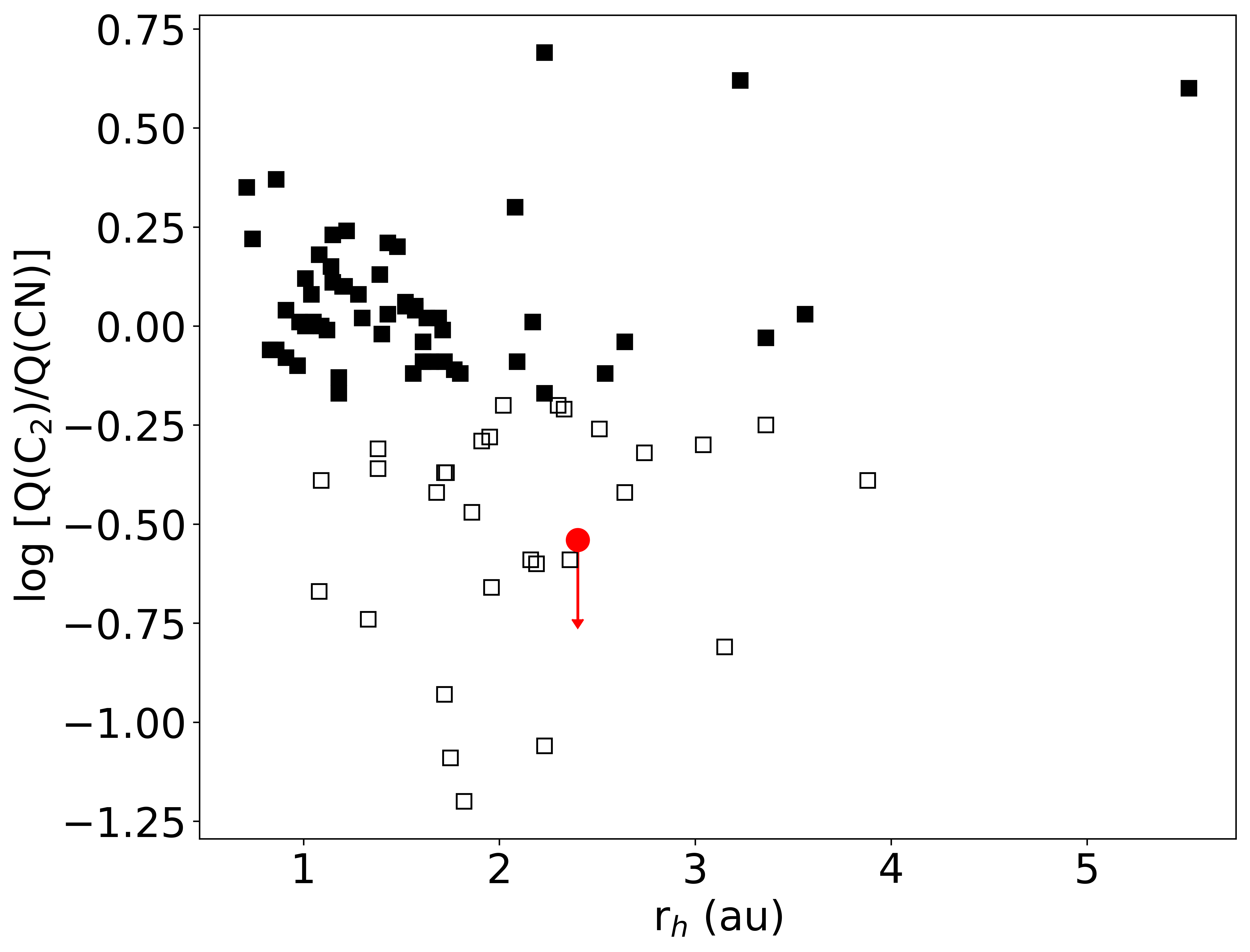}
			\includegraphics[scale=0.38]{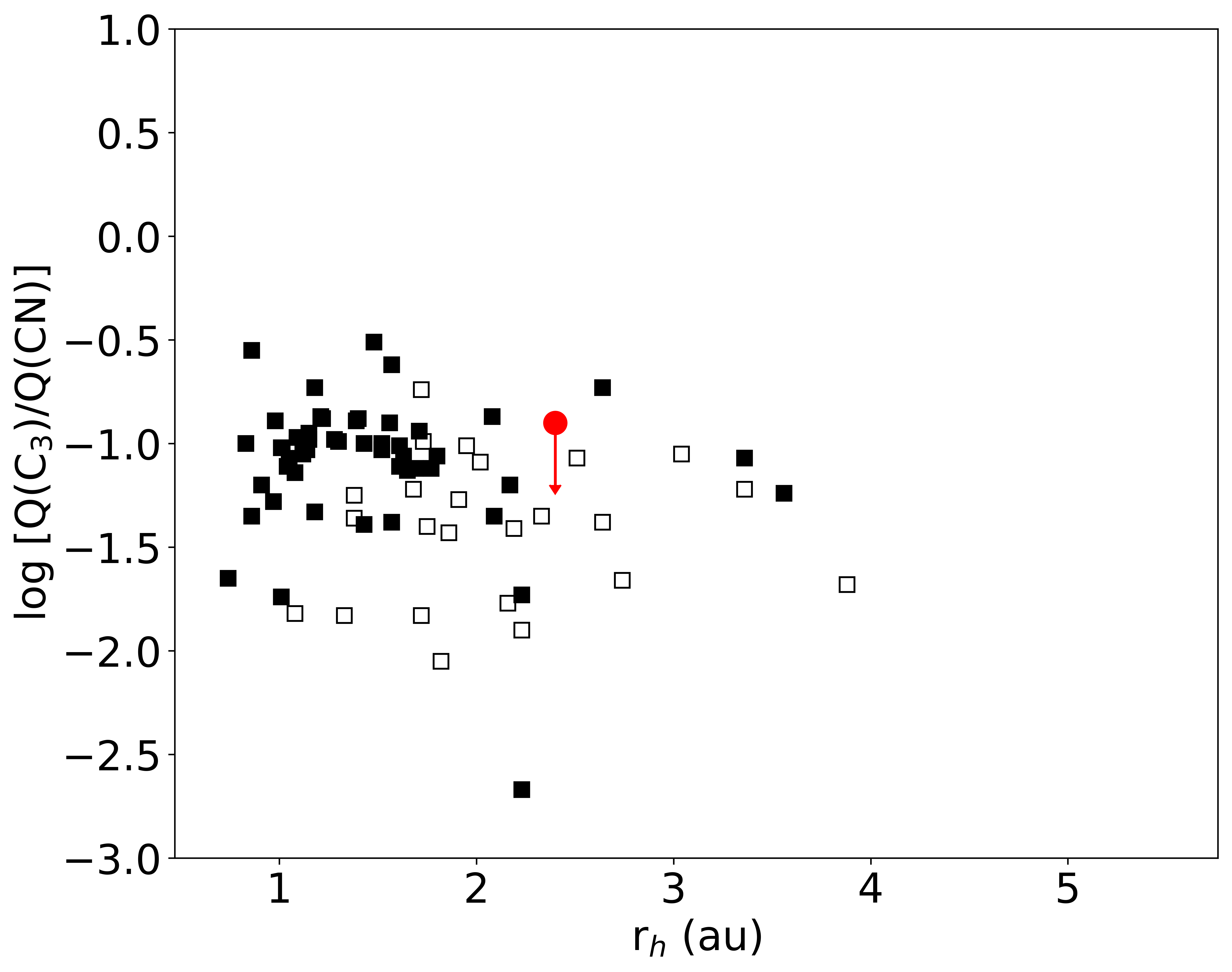}
			\caption{(a) Comparison between the upper limit for 2I (red data point) and Q(C$_2$)/Q(CN) from the \cite{osip2003} database for a variety of comets with different Q(CN). Full symbols represent typical comets and open symbols represent carbon-chain depleted comets as defined in \cite{ahearn1995}. Error bars are omitted for clarity.  (b) Same as (a) but for Q(C$_3$)/Q(CN).}
			\label{fig:CNabundances}
		\end{center}
	\end{figure*}
	
	The definition of carbon-chain depleted comets varies among studies. Some authors define them as being depleted in both $\mathrm{C_2}$ and $\mathrm{C_3}$ \citep{cochran2012} while others only consider the $\mathrm{C_2}$ abundance \citep{ahearn1995,fink2009}. Consequently, the proportion of depleted comets varies among studies, even though they are similar if only $\mathrm{C_2}$ is considered. Within the restricted data set of \cite{ahearn1995}, 29\% of the comets are depleted in $\mathrm{C_2}$ and 28\% according to \cite{fink2009}. \cite{cochran2012} have 9\% of comets in their restricted data set depleted in both $\mathrm{C_2}$ and $\mathrm{C_3}$, while 25\% are depleted only in $\mathrm{C_2}$. Carbon-chain-depleted comets therefore make up about 30\% of observed Solar System comets, with the majority of those being JFCs.
	
	Using the same scale lengths as those used here to compute gas production rates, \cite{ahearn1995} defined carbon-chain depleted comets as comets with log[Q($\mathrm{C_2}$)/Q(CN)]$<-0.18$. According to this definition, 2I is indeed depleted with log[(Q($\mathrm{C_2}$)/Q(CN)]$<-0.54$. In Fig. \ref{fig:CNabundances}, we compare the composition of 2I to the comets in the database of \cite{osip2003} (an online version of the database of \cite{ahearn1995}). On the left side we see that the upper limit of the Q($\mathrm{C_2}$)/Q(CN) measured for 2I is lower than most of the comets in that database, including the majority of carbon-chain depleted comets, and places 2I among the highly depleted comets. On the right side of Fig. \ref{fig:CNabundances}, we show a similar plot for the Q($\mathrm{C_3}$)/Q(CN) ratio. Our upper limit on the $\mathrm{C_3}$ production rate only indicates that the $\mathrm{C_3}$ abundance in the coma of 2I is not enhanced with respect to CN, as compared to the bulk of comets. 
	
	Studies have shown that the ratio between $\mathrm{C_2}$ and CN abundances changes with the heliocentric distance of the comet, and tends to decrease with increasing heliocentric distances above 1 au \citep{langland2001}. However, even when compared to typical comets observed at similar heliocentric distances, 2I appears severely depleted in $\mathrm{C_2}$, as shown in Fig. \ref{fig:CNabundances}. %From our CN production rates and Af$\rho$ measurements, we derived log(A(0)f$\rho$/Q(CN))=-22.18. This is consistent with the mean value of -22.61$\pm$0.32 reported by \citep{ahearn1995} for the carbon-chain depleted in their sample, while the mean value of the log(A(0)f$\rho$/Q(CN)) for typical comets if of -23.32$\pm$0.43. comets reported by \citep{ahearn1995} and above the mean value of typical comets of -23.32$\pm$0.43.
	
	Even though the origin of the carbon-chain depletion in Solar System comets is still debated, evidence suggest that it is intrinsic rather than evolutionary. Indeed, short-period comets observed at successive perihelion passages do not show significant changes in composition \citep{ahearn1995}. Moreover, observations of the split comet 73P/Schwassmann-Wachmann~3 in 2003 did not show any composition difference among the fragments, nor after the outburst, compared to previous passages of the comet \citep{schleicher2011}. If primordial, the composition difference between typical and depleted comets points to different formation regions for these comets. The $\mathrm{C_2}$ depletion of 2I might then indicate formation conditions similar to the carbon-chain depleted Solar System comets. On the other hand, the apparent very slow rise of 2I gas activity so far is reminiscent of the behaviour of dynamically new comets approaching the Sun for the first time \citep{ahearn1995}. Longer-term monitoring of the activity of 2I and actual detection of species other than CN, allowing  abundance ratios to be computed, are necessary to determine the likeness of 2I to the various families of Solar System comets. This is likely to be achieved in the coming weeks, while the comet becomes brighter and more easily observable to numerous telescopes.
	
	Finally, our non-detections of OH give log[Q(OH))/Q(CN)]$<3.0$. This upper limit is consistent with the abundance ratio of these species in both typical and depleted comets in \cite{ahearn1995}. Therefore, assuming OH derives as usual from the photodissociation of H$_2$O, the present data are consistent with a normal CN/OH ratio and there is no evidence for an abnormal water abundance in 2I.
	
	\section{Conclusions}
	\label{sec:conclusions}
	In this letter, we present spectroscopic observations of the first active interstellar object 2I/Borisov with the WHT and INT telescopes in la Palma, complemented by broadband and narrow-band imaging with the TRAPPIST-North telescope. We determine an upper limit of the $\mathrm{C_2}$ production rate and abundance relative to CN. We conclude that 2I is strongly depleted in $\mathrm{C_2}$, and would be classified as a carbon-chain depleted comet in our Solar System. Future spectroscopic observations should allow us to better constrain the extent of the $\mathrm{C_2}$ depletion in 2I and to further characterise the composition of its nuclear ices.

	{\it Acknowledgements}
	We thank Ovidiu Vaduvescu, Lilian Dominguez and Ian Skillen of the Isaac Newton Group for performing observations for us under service programme SW2019b04. The WHT and INT are operated on the island of La Palma by the Isaac Newton Group of Telescopes in the Spanish Observatorio del Roque de los Muchachos of the Instituto de Astrofísica de Canarias. TRAPPIST is a project funded by the Belgian Fonds (National) de la Recherche Scientifique (F.R.S.-FNRS) under grant FRFC 2.5.594.09.F. TRAPPIST-North is a project funded by the University of Liege, in collaboration with Cadi Ayyad University of Marrakech (Morocco). E.J is F.R.S.-FNRS Senior Research Associate. AF and CS acknowledge support for this work from UK STFC grants ST/P000304/1 and ST/L004569/1. KJM, JTK, and JVK acknowledge support through awards from NASA 80NSSC18K0853. 
	
	\bibliographystyle{aa}
	\bibliography{2I}
	
	%\begin{appendix}
	%\section{Dust colors and Af$\rho$}
	
	%\end{appendix}
	

\end{document}